\documentclass[12pt, a4paper]{article}
\usepackage[margin=1in]{geometry}
\usepackage[utf8]{inputenc}
\usepackage{latexsym}
\usepackage{changepage}  
\usepackage{hyperref}
\usepackage{csvsimple}
\usepackage{titlesec}
\titlelabel{\thetitle.\quad}

\makeatletter
\renewcommand{\@seccntformat}[1]{~\csname the#1\endcsname{}.}
\makeatother

\newcommand{\Section}{\section}

\makeatletter
\csvset{
  my hack/.style={
    late after line=\\\hline,
    late after last line=\csv@tablefoot\end{tabular}\csv@posttable},
}
\makeatother

\begin{document}

\begin{center}
    
    \Large\textbf{ Identifying a 3-vertex strongly biconnected directed subgraph with minimum number of edges}   \\[18pt]
    
    \normalsize{Azzam Habib}\\
    
       \normalsize{\noindent Faculty of Informatics Engineering, Tishreen University, Syria}\\
    
    \rule{\textwidth}{0.6pt}
    \begin{abstract}
A strongly connected graph is strongly biconnected if after ignoring the direction of its edges we have an undirected graph with no articulation points. 
A 3-vertex strongly biconnected graph is a strongly biconnected digraph that has the property that deleting any two vertices in this graph leaves 
a strongly binconnected subgraph. Jaberi \cite{RaedJaberi72021} presented approximation algorithms for minimum cardinality 2-vertex strongly biconnected directed subgraph problem.
We will focus in this paper on polynomial time algorithms which we have implemented for
producing spanning subgraphs that are 3-vertex strongly biconnected.
    \end{abstract}
    \textbf{\normalsize {keywords}\\}
    \small{ Graph algorithms, Approximation algorithms, Connectivity}
    \rule{\textwidth}{0.4pt}
\end{center} 
\Section{Introduction}
 A new kind of directed graphs, called strongly biconnected directed graphs was mentioned by Wu and Grumbach \cite{ZhilinWuStephaneGrumbach}. A strongly connected graph is strongly biconnected if after ignoring the direction of its edges we have an undirected graph with no articulation points. 
A 3-vertex strongly biconnected graph is a strongly biconnected directed graph that has the property that deleting any two vertices in this graph leaves 
a strongly binconnected subgraph. 
Testing 2-vertex connectivity in directed graphs \cite{LoukasGeorgiadis2010,GiuseppeItalianoLuigiLauraFedericoSantaroni,DonatellFirmanietal} can be done in linear time. Additionally, testing 2-connectivity in undirected graphs \cite{RobertEndreTarjan,JensSchmidt2013,Gabow2000,SandersMehlhornDietzfelbingerDementiev0192019,Cormen2022} can be done in linear time. 
Calculating a k-vertex connected spanning subgraph with minimum number of edges is NP-hard Problem \cite{GareyDavidJohnson,JosephCheriyanRamakrishnaThurimella}. 
Furthermore, Identifying a k-vertex-strongly biconnected spanning subgraph with minimum number of edges is also NP-hard \cite{RaedJaberi72021}. 
Cheriyan and Thurimella \cite{JosephCheriyanRamakrishnaThurimella} provided approximation algorithms for finding a k-vertex connected spanning subgraph with minimum number of edges. In addition,
Georgiadis et al. \cite{LoukasGeorgiadisGiuseppeItalianoAikateriniKaranasiou,LoukasGeorgiadis} gave
improved versions of these algorithms and efficient algorithms for k = 2. The 2-vertex connected directed
graphs have the property that they do not have any strong articulation point. Strong articulation
points can be found in linear time  by means of  algorithms provided by Italiano et al. \cite{GiuseppeItalianoLuigiLauraFedericoSantaroni,DonatellFirmanietal}.
By results \cite{JEdmonds1972,1985WMader}, each minimal 2-vertex connected directed graph has at most 4n edges. Jaberi \cite{RaedJaberi72021} presented approximation algorithms for minimum cardinality 2-vertex strongly biconnected directed subgraph problem. Moreover, he studied minimum cardinality 2-edge strongly biconnected directed subgraph problem \cite{RaedJaberi2022}.
Here we study the minimum cardinality 3-vertex strongly biconnected spanning directed subgraph problem.

\section{Finding $3$-vertex strongly biconnected  subgraph}
In this section we  describe  approximation algorithms for 3-vertex strongly biconnected graph.
Let $G = (V, E)$ be a 3-vertex strongly biconnected graph, we say that G is minimal if $G \setminus \lbrace e \rbrace$ is not 3-vertex strongly biconnected for any edge $e\in E$.\\
In order to identify a minimal spanning subgraph that is 3-vertex strongly biconnected, we used the simplest approach, just by removing edges one by one and check if the remaining directed  subgraph is 3-vertex strongly biconnected. Algorithm 1 gives the pseudocode of this simple approach which runs in $O(n^{2}m(n + m))$ time.\\\\
\rule{17cm}{0.2mm}\\
\textbf{Algorithm 1}\\
\rule{17cm}{0.2mm}\\
\textbf{Input}: A 3-vertex strongly biconnected directed graph $G=(V,E)$\\
\textbf{Output}: A minimal $3$-vertex strongly biconnected  subgraph of $G$\\
1. \textbf{for} each edge e in G \textbf{do}\\
2.\space \space \space\space \textbf{if} the subgraph after deleting $(i,j)$ from G is 3-vertex strongly biconnected \textbf{then} \\
3.\space \space \space\space \space\space remove the directed edge $(i,j)$ from the directed graph $G$\\
4.\space Output $G$\\
\rule{17cm}{0.2mm}\\\\

Jaberi \cite{RaedJaberi72021} showed that every minimal 2-vertex strongly biconnected spanning subgraph has at most 7n edges. His proof is based on the results \cite{JEdmonds1972,1985WMader,1971WMader,1972WMader}. Using the results \cite{JEdmonds1972,1985WMader,1971WMader,1972WMader} we can show that each minimal 3-vertex-strongly biconnected directed graph has 10n edges. Consequently, Algorithm 1 has approximation factor of 10/3.

 Cheriyan and Thurimella \cite{JosephCheriyanRamakrishnaThurimella} gave  $(1+1/k)$  approximation algorithms for the minimum cardinality $k$-vertex-connected spanning subgraph problem. For k=2, the subgraph returned by this algorithm is 2-vertex-connected but not a solution for minimum cardinality 2-vertex strongly biconnected directed subgraph problem. Jaberi \cite{RaedJaberi72021} described a modified version of this algorithm which is able to obtain a $2$-vertex-strongly biconnected subgraph. We will introduce a modified version of these algorithms which can produce a $3$-vertex-strongly biconnected subgraph. Algorithm 2 gives the pseoducode of this approach which consist of two steps.\\
step1: identify a $2$-vertex-strongly biconnected subgraph $G^{+} = (V, E^{+})$ using first algorithm presented by Jaberi \cite{RaedJaberi72021}.\\
step2: run the MINIMAL algorithm but only for the edges in $E \setminus E^{+}$\\
\rule{17cm}{0.2mm}\\
\textbf{Algorithm 2}\\
\rule{17cm}{0.2mm}\\
\textbf{Input}: A 3-vertex strongly biconnected directed graph $G=(V,E)$\\
\textbf{Output}: A $3$-vertex strongly biconnected spanning subgraph of $G$\\
1.  find 2-vertex strongly biconnected spanning subgraph of $G$, $G^{+}=(V,E^{+})$\\
2. \textbf{for} each edge $e\in E \setminus E^{+}$ \textbf{do}\\
3.\space \space \space\space \textbf{if} $G\setminus \lbrace e\rbrace$ is 3-vertex strongly biconnected \textbf{then} \\
4.\space \space \space\space \space\space remove $e$ from $G$\\
5.\space Output $G$\\
\rule{17cm}{0.2mm}\\\\
 \section{The results of our experiments}   
Here we report the results of the experiments that we conducted. 

\begin{center}
\begin{tabular}{|c||c||c||c||c|}
   \hline 
   Input & Algorithm1 & Algorithm1 & Algorithm2 & Algorithm2 \\ 
   \hline 
   ( V , E ) & Time & Edges & Time & Edges \\ 
   \hline 
   ( 10 , 80 ) &2 s & 32 & 1 s & 33  \\ 
   \hline 
   ( 20 , 160 ) &19 s & 67 & 14 s & 67  \\ 
   \hline 
   ( 30 , 240 ) &1 m 13  s & 95 & 44 s & 101  \\ 
   \hline 
   ( 40 , 420 ) &3 m 4  s  & 130 & 2 m 22 s & 135  \\ 
   \hline 
   ( 50 , 417 ) & 5 m  36 s & 163 & 3 m 34 s & 169 \\ 
   \hline 
   ( 60 , 580 ) & 11 m  46 s & 189 & 8 m 51 s & 196 \\ 
   \hline 
   ( 70 , 560 ) & 14 m  54 s & 229 & 10 m 8 s & 234 \\ 
   \hline 
   ( 80 , 740 ) & 26 m  16 s & 257 & 17 m 48 s & 270 \\ 
   \hline 
   ( 90 , 720 ) & 30 m  34 s & 291 & 20 m 10 s & 298 \\ 
   \hline 
   ( 100 , 819 ) & 50 m 28 s & 333 & 28 m 3 s & 346 \\ 
   \hline 
   ( 120 , 1160 ) & 1 h 29 m 28 s & 391& 1 h 5 m 6 s & 401\\ 
   \hline 
   \end{tabular}
\end{center}

We implemented all our algorithms in Java without the use of any external graph library. The experiments were conducted on a 64-bit Intel(R)/Windows machine running Windows 21H1, with a 1.99 GHz Intel i7- 8550U, 12 GB of RAM, 8 MB of L3 cache, and each core has a 256 KB private L2 cache. All experiments were executed on a single core without using any parallelization. We report CPU times measured with the nanoTime function.
For our experimental study, we used a collection of random generated graphs, we constructed 3-vertex strongly biconnected graphs as follows. We generated random graphs with n node and 8n random edges, then we keep adding edges one by one until the graph become 3-vertex strongly biconnected. For these graphs, notice that the output of the algorithms has less than $4n$ edges. Moreover, the number of edges returned by Algorithm1  is less than the number of edges produced by Algorithm 2.


\addcontentsline{toc}{section}{References}
\begin{thebibliography}{1}
\bibitem{Cormen2022}Thomas H. Cormen, Charles E. Leiserson, Ronald L. Rivest, Clifford Stein:
Introduction to Algorithms, fourth Edition. MIT Press, 2022, ISBN 9780262046305
\bibitem{JosephCheriyanRamakrishnaThurimella} Joseph Cheriyan, Ramakrishna Thurimella:
Approximating Minimum-Size k-Connected Spanning Subgraphs via Matching. SIAM J. Comput. 30(2): 528-560 (2000)
	\bibitem{JEdmonds1972}J. Edmonds: Edge-Disjoint Branchings, In Rustin, R., editor, Combinatorial Algorithms, pp. 91-96, Academic Press (1973)
\bibitem{DonatellFirmanietal} Donatella Firmani, Loukas Georgiadis, Giuseppe F. Italiano, Luigi Laura, Federico Santaroni:
Strong Articulation Points and Strong Bridges in Large Scale Graphs. Algorithmica 74(3): 1123-1147 (2016)
\bibitem{Gabow2000}H. N. Gabow. Path-based depth-first search for strong and biconnected components.
Information Processing Letters, 74(3–4):107–114, 2000.
\bibitem{LoukasGeorgiadis2010} Loukas Georgiadis:
Testing 2-Vertex Connectivity and Computing Pairs of Vertex-Disjoint s-t Paths in Digraphs. ICALP (1) 2010: 738-749
	\bibitem{LoukasGeorgiadis}Loukas Georgiadis:
Approximating the Smallest 2-Vertex Connected Spanning Subgraph of a Directed Graph. ESA 2011: 13-24
	\bibitem{GareyDavidJohnson} M. R. Garey, David S. Johnson:
Computers and Intractability: A Guide to the Theory of NP-Completeness. W. H. Freeman 1979, ISBN 0-7167-1044-7
	\bibitem{LoukasGeorgiadisGiuseppeItalianoAikateriniKaranasiou}Loukas Georgiadis, Giuseppe F. Italiano, Aikaterini Karanasiou:
Approximating the smallest 2-vertex connected spanning subgraph of a directed graph. Theor. Comput. Sci. 807: 185-200 (2020)
	\bibitem {GiuseppeItalianoLuigiLauraFedericoSantaroni}Giuseppe F. Italiano, Luigi Laura, Federico Santaroni:
Finding strong bridges and strong articulation points in linear time. Theoretical Computer Science 447: 74-84 (2012)
	\bibitem{RaedJaberi72021}Raed Jaberi:
Minimum 2-vertex strongly biconnected spanning directed subgraph problem, Discrete Mathematics Letters 7 (2021) 40-43
DOI: 10.47443/dml.2021.0024

\bibitem{RaedJaberi2022} Raed Jaberi:
Minimum 2-edge strongly biconnected spanning directed subgraph problem. CoRR abs/2207.03401 (2022)
\bibitem{1972WMader} W. Mader: Ecken vom Grad n in minimalen n-fach zusammenhängenden Graphen. Archiv der Mathematik, 23:219-224,(1972)
	
	\bibitem{1971WMader} W. Mader, Minimale n-fach kantenzusammenhängende Graphen. Mathematische Annalen Volume: 191, page 21-28(1971) 
		\bibitem{1985WMader} W. Mader,
	Minimal $n$-fach zusammenhängende Digraphen. Journal of Combinatorial Theory, Series B
Volume 38, Issue 2, April 1985, Pages 102-117 	
	\bibitem{JensSchmidt2013} Jens M. Schmidt:
A simple test on 2-vertex- and 2-edge-connectivity. Inf. Process. Lett. 113(7): 241-244 (2013)
	
\bibitem{SandersMehlhornDietzfelbingerDementiev0192019}Peter Sanders, Kurt Mehlhorn, Martin Dietzfelbinger, Roman Dementiev:
Sequential and Parallel Algorithms and Data Structures - The Basic Toolbox. Springer 2019, ISBN 978-3-030-25208-3, pp. 1-434	
	\bibitem{RobertEndreTarjan}
	Robert Endre Tarjan:
Depth-First Search and Linear Graph Algorithms. SIAM Journal on Computing,  1(2): 146-160 (1972)
	\bibitem{ZhilinWuStephaneGrumbach} Zhilin Wu, Stéphane Grumbach,
Feasibility of motion planning on acyclic and strongly connected directed graphs. Discrete Applied Mathematics 158(9): 1017-1028 (2010)




\end{thebibliography}
\end{document}